\documentclass[a4paper]{article}

\usepackage{amsthm,amsmath,amssymb,amsfonts,mathrsfs,dsfont}
\usepackage{graphicx}
\usepackage{bm}
\usepackage{geometry}

\newtheorem{theorem}{Theorem}

\newcommand{\diff}[2]{\frac{\partial{#1}}{\partial{#2}}}
\newcommand{\Alg}[0]{\mathscr{A}}

\begin{document}

	{\title{\protect\vspace*{-2cm}Symmetry nonintegrability\\ for~extended $K(m,n,p)$ equation}}
	\author{Jakub Va\v{s}\'\i{}\v{c}ek\\ Silesian University in Opava, Mathematical Institute,\\
Na Rybn\'\i{}\v{c}ku 1, 746 01 Opava, Czech Republic\\ E-mail: {\tt jakub.vasicek@math.slu.cz}}
		
\maketitle
	\begin{abstract}
		\indent 
In the present paper we study symmetries of extended $K(m,n,p)$ equations and prove that the equations from this class have no generalized symmetries of order greater than five and hence are not symmetry integrable.
	\end{abstract}

\section{Introduction}
Consider the equation, cf.\ \cite{dey98,vod13} and references therein:
\begin{equation}\label{ekmnp}
 u_t=a(u^p)_{xxxxx}+b(u^n)_{xxx} + c(u^m)_{x} + f(u),
\end{equation}
to which we shall refer to as the {\em extended $K(m,n,p)$ equation},  
as for $f=0$ equation \eqref{ekmnp} boils down to the $K(m,n,p)$ equation from \cite{dey98}:
\begin{equation}
\label{deyEq}
u_t=a(u^p)_{xxxxx}+b(u^n)_{xxx} + c(u^m)_{x}.
\end{equation}
Here $a,b,c$ are real constants and $m,n,p$ are integers.

Equation \eqref{deyEq}, as well as its special case for  $a=0$, the well-known $K(m,n)$ equation, see e.g.\ \cite{ros93, vod13} and references therein, are the subject of intense research since they admit compacton solutions \cite{dey98,ros93, ros18}. Such solutions are of significant interest since they share many features with the famous solitons but, unlike the latter, have compact support \cite{ros93,ros18}.

It is well known that integrable systems have rich symmetry algebras, cf.\ e.g.\ \cite{mos, olv93, s18} and references therein. A simple observation that the computation of symmetries is, to a large extent, an algorithmic procedure has, in a series of nontrivial developments, lead to symmetry-based integrability tests and the related notion of symmetry integrability \cite{mos, olv93, mik87, mik09, s05, sev16}. For a recent survey on the latter, see \cite{mik09}, and for integrability in general
see e.g.\ \cite{s18, mik87, mik09, kra11} and references therein.\looseness=-1

Recall that integrable systems enjoy many remarkable properties like infinite hierarchies of conservation laws or large sets of explicit exact solutions, cf.\ e.g.\ \cite{mik87, mik09, olv93, s05, s18} and references therein, which makes clear the importance of establishing integrability of a given equation.
 
In the remaining part of the present paper we show that for $a\neq 0$ and $p\neq 1,-4$ equation (\ref{ekmnp}), and hence the original $K(m,n,p)$ equation (\ref{deyEq}) as well, is not symmetry integrable as it has no generalized symmetries of order greater than five, see Theorem~\ref{t1} below for details. 

\section{Main result}
For partial derivatives in $x$ we employ the usual notation $u_{j}=\partial^j u/\partial x^j$, so $u_{0}\equiv u$, $\partial u/\partial x=u_{1}\equiv u_x$, $\partial^2 u/\partial x^2\equiv u_{xx}=u_{2}$, etc.

Let $\Alg$ be an algebra of smooth functions of $x,t,u^{1/5},u^{-1/5}$ and finitely many $u_{j}$ (we stress that of course different functions from  $\Alg$ may depend on a different, but always finite, number of $u_j$). 
In spirit of \cite{olv93} the elements of $\Alg$ will be referred to as \emph{differential functions}, and it will be tacitly assumed from now on that all functions encountered below are differential functions unless otherwise explicitly stated.

\begin{theorem}\label{t1}
For any $f(u)$ and arbitrary real constants $a,b,c$ and arbitrary integers $m,n,p$ such that $a\neq 0$ and $p\neq 1,-4$, the extended $K(m,n,p)$ equation
 \begin{equation}
  \label{tf-eq}
 u_t=a(u^p)_{xxxxx}+b(u^n)_{xxx} + c(u^m)_{x} + f(u)
 \end{equation}
has no generalized symmetries of order greater than $5$.
\end{theorem}

\section{Preliminaries}
Here we briefly recall a few known results following mostly \cite{mik87, mik09, olv93, s99,  sev16}.

Consider for now an evolution equation in one dependent and two independent variables of the form
\begin{equation}
\label{evo}
u_t=F(x,u,u_1,\dots,u_{n}), \qquad n \geq 2.
\end{equation}

The operators of total derivatives $D_x$ and $D_t$ 
then take the form \cite{mik87,mik09} \begin{equation} D_{x}= \frac{\partial }{\partial x} + \sum^\infty_{i=0}u_{i+1}\frac{\partial }{\partial u_i},\quad D_{t}= \frac{\partial }{\partial t} + \sum^\infty_{i=0}D_x^i(F)\frac{\partial }{\partial u_i}\end{equation}

An evolutionary vector field $\mathbf{v}_Q=Q\partial/\partial u$ is called \cite{olv93} a \emph{generalized symmetry} for \eqref{evo} if its characteristic $Q$ satisfies \looseness-1
\begin{equation}
\label{chsym}
D_t(Q) =\mathrm{\textbf{D}}_F(Q).
\end{equation}
Here for an $F=F(x,t,u,u_1,\dots,u_{k})$ we denote \cite{olv93} by $\mathrm{\textbf{D}}_F$ its (formal) Fr\'echet derivative $$\mathrm{\textbf{D}}_F=\sum_{j=0}^{k}{\diff{F}{u_{j}} D_x^j}.$$

\section{Proof of the main result}
\label{section4}



To simplify writing, denote now by $F$ the right-hand side of (\ref{tf-eq}) and let $n=\mathrm{ord}\, F$, so in our case $n=5$.

It follows from well-known general results, cf.\ e.g.\ \cite{olv93, s99}, that if $Q$ is a characteristic of generalized symmetry for (\ref{tf-eq}) and $k=\mathrm{ord}\,Q>1$ then we have
\[
\diff{Q}{u_k}=c_k(t)\left(\displaystyle{\diff{F}{u_n}}\right)^{k/n}
\]
for some function $c_k(t)$.

Moreover, Theorem 1 from \cite{s99} then implies that if furthermore $k>n$ then we have
\begin{equation}\label{c}
D_x\left(\displaystyle\diff{Q}{u_{k-n+1}}  (\rho_{-1})^{n-k-1} \right) = D_x(h_{k-n+1}) +  \displaystyle\frac{k}{n^2}c_k(t)D_t \left(\rho_{-1} \right) + \displaystyle\frac{1}{n}\dot{c}_k(t)\rho_{-1},
\end{equation}
for some differential function $h_{k-n+1}$.

Here $\rho_{-1}$ stands for the so-called minus first canonical density \cite{mik87}:
$$\rho_{-1}=\left(\diff{F}{u_n}\right)^{-1/n}.$$

It is immediate that for (\ref{c}) to hold, the expression
\[
\displaystyle\frac{k}{n^2}c_k(t)D_t\left( \left(\displaystyle\diff{F}{u_n}\right)^{-1/n}\right) + \displaystyle\frac{1}{n}\dot{c}_k(t)\left(\displaystyle\diff{F}{u_n}\right)^{-1/n}
\]
should belong to the image of $D_x$.

In turn, a well-known, see e.g.\ \cite{olv93}, necessary condition for the latter is
\begin{equation}\label{co}\frac{\delta}{\delta u}\left( \displaystyle\frac{k}{n^2}c_k(t)D_t\left( \left(\displaystyle\diff{F}{u_n}\right)^{-1/n}\right) + \displaystyle\frac{1}{n}\dot{c}_k(t)\left(\displaystyle\diff{F}{u_n}\right)^{-1/n}\right)=0,\end{equation}
where by $\displaystyle\frac{\delta }{\delta u}$ we mean the operator of variational derivative, cf.\ e.g.\ \cite{olv93,sev16}, $$\frac{\delta }{\delta u}=\sum_{i=0}^\infty (-1)^iD_x^i \circ \frac{\partial }{\partial u_i}.$$

Denote the left hand-side of \eqref{co} by $C$ and observe that
\begin{equation}\label{co1}\diff{C}{u_{4}}= \frac{2k}{625}(ap)^{4/5}c_ku^{(4/5)p-19/5}u_x(p-1)(p+4)(3p+2).
\end{equation}

Clearly, for $C$ to vanish it is necessary that $\partial{C}/\partial{u_{4}}$ vanishes as well.

It is obvious from \eqref{co1} that under the assumptions of our theorem $\partial{C}/\partial{u_{4}}$ cannot vanish unless $c_k=0$, and thus under the said assumptions equation (\ref{tf-eq}) cannot have a generalized symmetry of order greater than $5$.



\section{Conclusion}
Thus, for $a\neq 0$ and $p\neq 1,-4$ equation \eqref{tf-eq} has no generalized symmetries of order greater than five and  cannot have an infinite hierarchy of generalized symmetries of arbitrarily high order, so \eqref{tf-eq} for $a\neq 0$ and $p\neq 1,-4$ is not symmetry integrable. This result paves the way to finding all generalized symmetries for \eqref{tf-eq}, or its special case \eqref{deyEq}, because it suffices now to find the generalized symmetries with the characteristics of order up to 5, which can be done for instance using computer algebra software like \cite{jets}. However, in doing so one runs into plenty of various special cases, treating all of which in detail would go beyond the scope of the present paper.\looseness=-1

	
\subsection*{Acknowledgments}

This research was supported by the Specific Research grant SGS/6/2017 of the Silesian University in Opava.

I would like to express my most sincere gratitude to Artur Sergyeyev for stimulating discussions and valuable comments.

A large part of the computations in the article were performed using the package \emph{Jets} \cite{jets}.
 \nopagebreak[4]

\end{document}